\documentclass[preprint,5p,times,twocolumn]{elsarticle}

\usepackage{amssymb}
\usepackage{amsmath}
 \usepackage{lineno}

\journal{Nuclear Instruments and Methods in Physics Research Section A}

\begin{document}

%%\linenumbers

\begin{frontmatter}

\title{A New Beam Monitor at NFS/SPIRAL2 Based on Position-Sensitive PPACs Detecting Fission Fragments from ${}^{238}$U$(n,f)$}

\author[ganil]{D. Ramos}
\author[ganil]{X. Ledoux}
\author[ijclab]{L. Audouin}
\author[ganil]{G. Fremont}
\author[ganil]{P. Gangnant}
\author[ganil]{J.~C. Foy}
\author[ijclab]{C. Le Naour}
\author[ijclab]{M. Maloubier}

\affiliation[ganil]{organization={GANIL, CEA/DRF/IRFU-CNRS/IN2P3},
             addressline={Boulevard Henri Becquerel, BP 55027},
             city={Caen Cedex 5},
             postcode={F-14076},
             country={France}}

\affiliation[ijclab]{organization={IJCLab, CNRS/IN2P3-Universit{\'e} Paris-Saclay},%Department and Organization
            addressline={15 rue Georges Clemenceau}, 
            city={Orsay},
            postcode={F-91400}, 
            country={France}}
\begin{abstract}
%% Text of abstract

A new experimental setup has been installed at the Time-Of-Flight area of the Neutrons For Science facility (NFS) at GANIL/SPIRAL2 for neutron beam monitoring. This setup consists of an array of Position-Sensitive Parallel-Plate Avalanche Counters (PS-PPACs) that detects both fission fragments in coincidence from secondary neutron-induced fission reactions in several ${}^{238}$U targets. The neutron energy is determined on an event-by-event basis using the Time-of-Flight method, and the reaction point within the U targets is reconstructed, enabling the measurement of the neutron beam flux and beam profile. The high transparency of the setup allows it to operate in parallel with other experiments running at NFS, thus providing an in-beam monitor of the neutron intensity. In this work, we report on the characteristics of this new setup, its operating principle, and the first results obtained using the high-intensity white-spectrum neutron beam at NFS. This beam is produced via reactions between a primary 40-MeV deuteron beam, accelerated in the SPIRAL2 LINAC, and a 8 mm-thick rotating beryllium converter target.    

\end{abstract}

\end{frontmatter}

%% \linenumbers

%% main text
\section{Introduction}
\label{sec_introduction}

SPIRAL2 (Syst{\`e}me de Production d’Ions Radioactifs Acc{\'e}l{\'e}r{\'e}s en Ligne – 2) is a cutting-edge nuclear physics facility located at GANIL (Grand Acc{\'e}l{\'e}rateur National d'Ions Lourds) in Caen, France. It is based on a new superconducting linear accelerator (LINAC), designed to deliver high-intensity beams, including deuterons, protons, and heavy ions~\cite{Ganil,Spiral2}.

The Neutrons For Science facility (NFS) is one of the experimental halls directly fed by the LINAC and is dedicated to the production of intense neutron beams and the study of neutron-related physics. NFS consists of two main sections: a converter room, where neutrons are produced from reactions between the primary proton or deuteron beam from the LINAC and $^{7}$Li or $^{9}$Be converter targets, and a time-of-flight (ToF) area. The ToF area is a 28-meter-long experimental hall separated from the converter room by a 3-meter-thick concrete wall. A collimator embedded in this wall defines the neutron beam that enters the ToF hall. The length of the ToF area allows multiple experiments to be conducted simultaneously at different positions along the flight path. The neutron beam at NFS can be either monoenergetic or continuous with energies ranging from 1 MeV to 40 MeV, depending on the selection of the primary beam and the thickness of the converter target. More details of the NFS facility are available in Refs.~\cite{NFS1,NFS2,NFS3}.

This manuscript reports on a new beam-monitoring setup installed at the NFS ToF area. The neutron beam is measured through secondary fission reactions $^{238}$U$(n,f)$ by detecting the outgoing fission fragments in an ensemble of position-sensitive Parallel-Plate Avalanche Counters (PS-PPACs). The 1-MeV threshold of the $^{238}$U(n,f) cross section makes this reaction suitable for studying the neutron-energy range of NFS, while also reducing the background from thermalized neutrons. 

The flux and profile of the white-energy-spectrum neutron beam, produced by the interaction of a 7.6-$\mu$A intensity, 440-kHz frecuency, and 40-MeV energy primary deuteron beam with the 8 mm-thick rotating $^{9}$Be converter target, were measured using this setup.

\begin{figure}[!]
\includegraphics[width=0.49\textwidth]{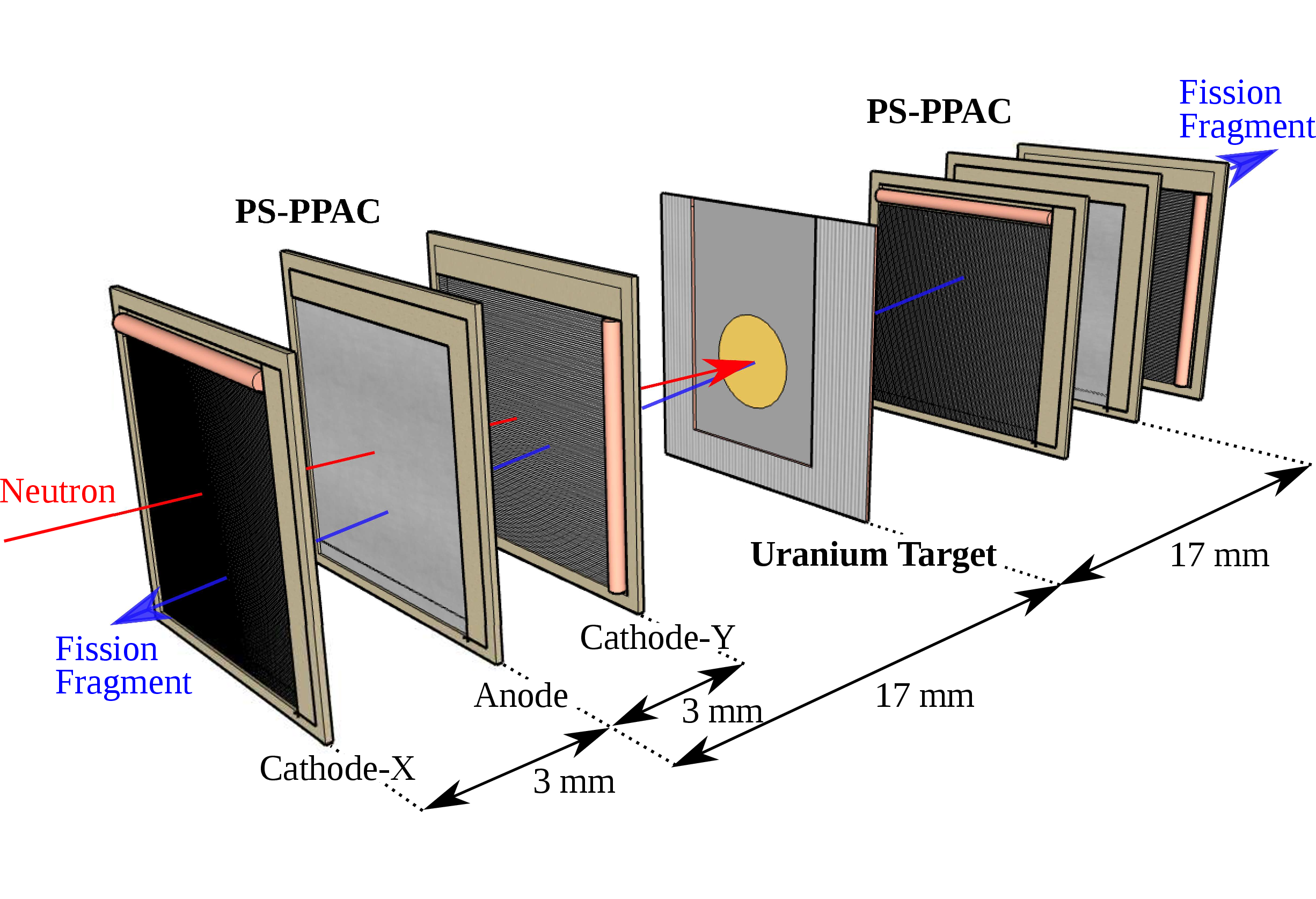}
\caption{Schematic view of the experimental setup. The Uranium target is surrounded by two PS-PPACs, meant for fission-fragments coincidence detection.}
\label{fig_fission}
\end{figure}

\section{Experimental Setup}
\label{sec_setup}

Two 298 ${\mu}g/cm^2$-thick and 8 cm-diameter $^{238}$U targets were placed in the neutron-beam line at 740.6 cm and 744.0 cm from the primary beryllium converter target. The uranium samples were electrodeposited on 2 $\mu$m-thick aluminium backings by molecular plating. Both fragments emitted from the fission of $^{238}$U were detected, in coincidence, with two position-sensitive Parallel-Plate Avalanche Counters (PS-PPACs) surrounding each target and separated from them by 17 mm, all positioned perpendicular to the beam axis. The reaction point in the $^{238}$U target is reconstructed by tracking the positions of both coincident fission fragments. A similar setup was recently used in the commissioning of the new nTOF spallation target, as reported in Refs.~\cite{Pavon,Spelta}.

The PS-PPACs detectors~\cite{ppacs1,ppacs2,ppacs3} are composed of three 1.5 $\mu$m-thick aluminized Mylar foils of 20x20 cm$^2$ of active area, separated by a 3 mm gap. The central foil is aluminized on both sides and acting as the anode. The two cathodes are aluminized on one side with 2 mm strips and 100 $\mu$m interstrip spacing. The strips are connected to a delay line of 3.6 ns/strip providing a position measurement by time difference between the two output signals from each edge of the delay line. The two strip cathodes are orthogonally placed to measure the positions (X,Y) of the fission fragment hit. A schematic representation of the setup is shown in Fig.~\ref{fig_fission}.

The angular converage of the setup ranges from 0 to 78 deg in polar angle, however, the system detection efficiency drops above 50 deg because the effective thickness of the target, backing, and foils increases, causing heavy fragments to be stopped before reaching the second cathode.

\begin{figure}[!]
\includegraphics[width=0.27\textwidth]{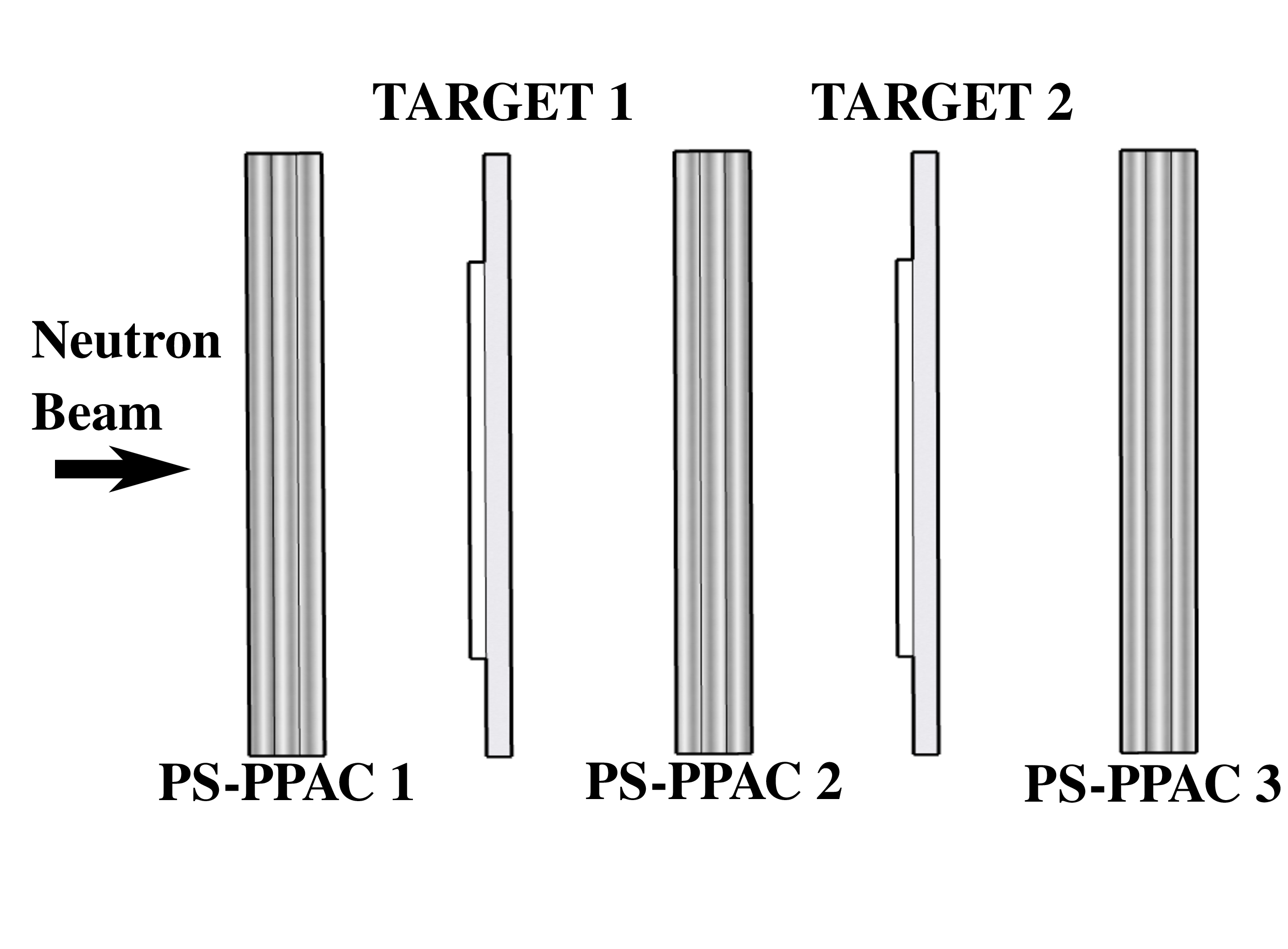}
\includegraphics[width=0.20\textwidth]{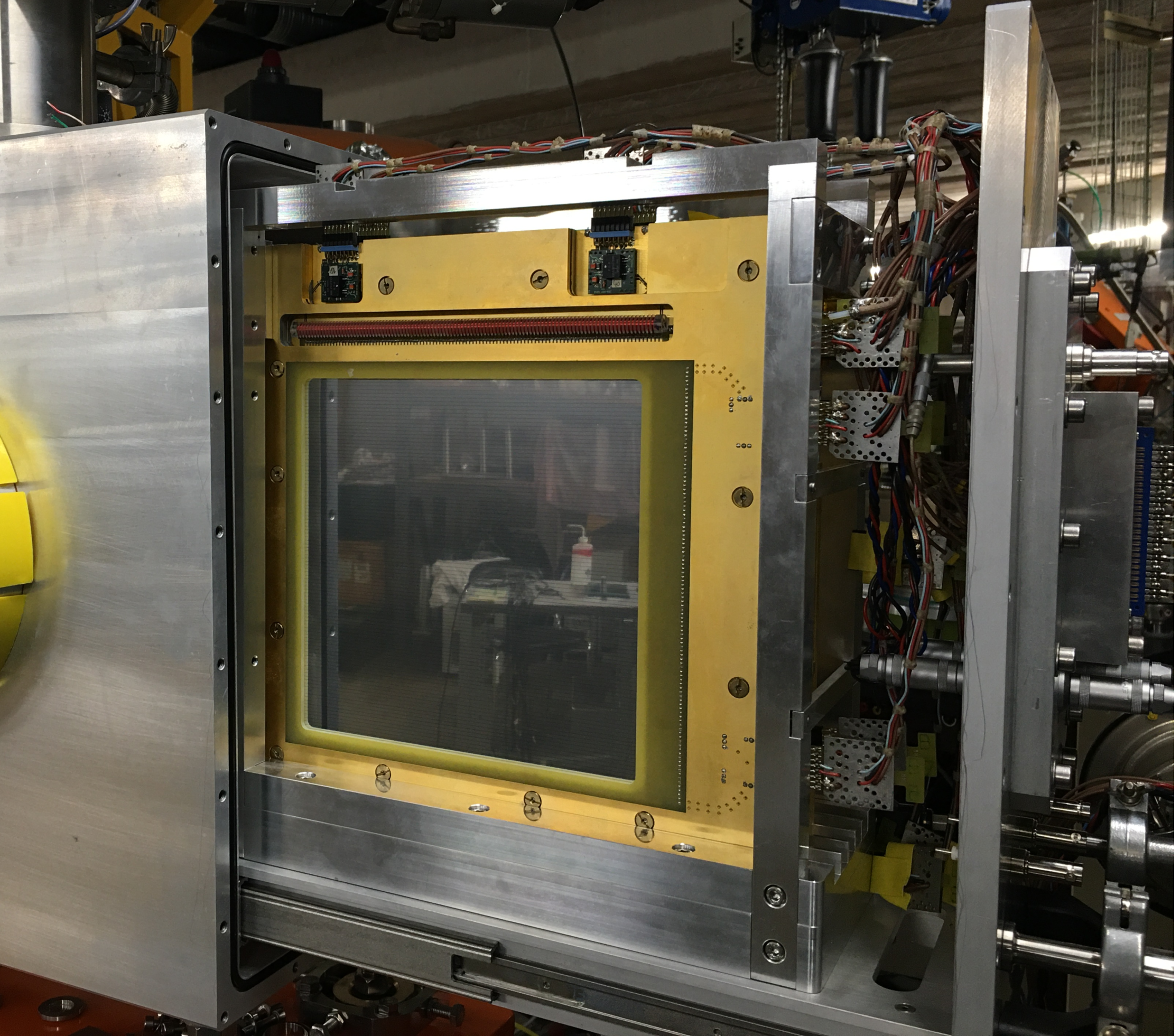}
\caption{Schematic layout of the detection system (left) and picture of the reaction chamber (right) allocating three PS-PPACs and two $^{238}$U targets.}
\label{fig_chamber}
\end{figure}

\begin{figure*}[t!]
\includegraphics[width=0.49\textwidth]{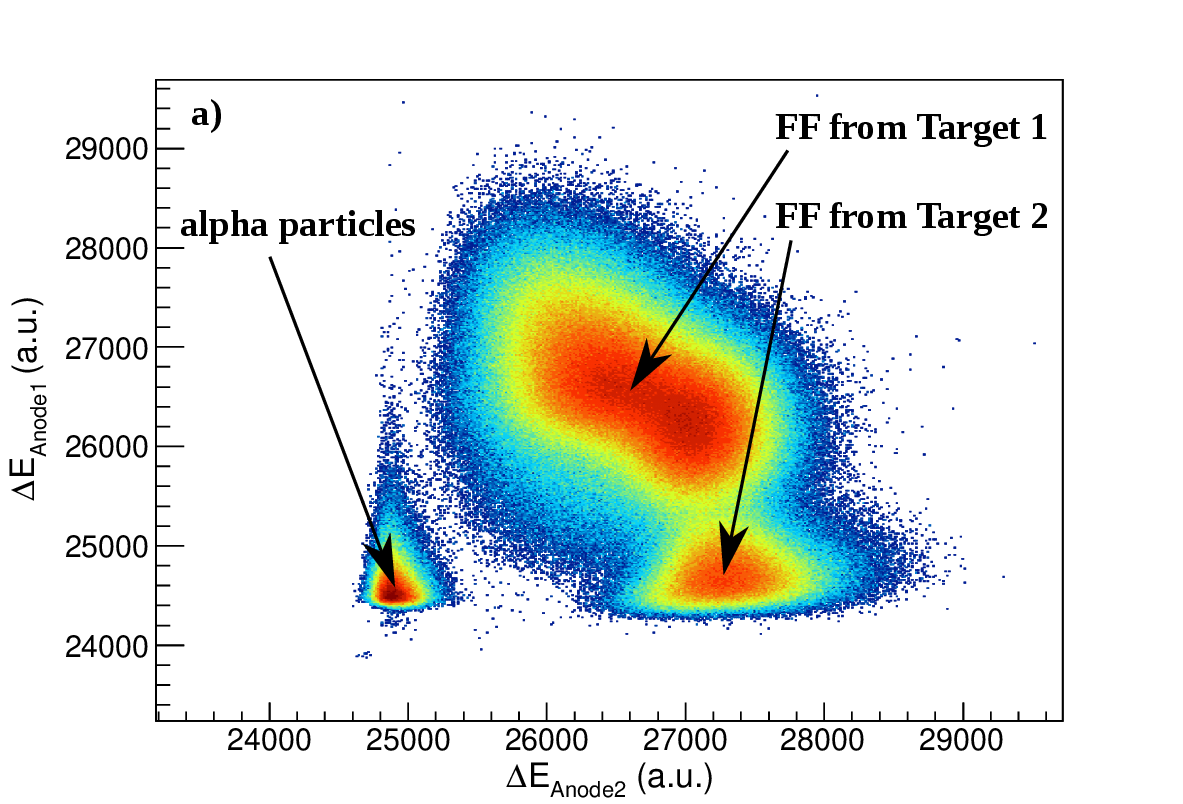}
\includegraphics[width=0.49\textwidth]{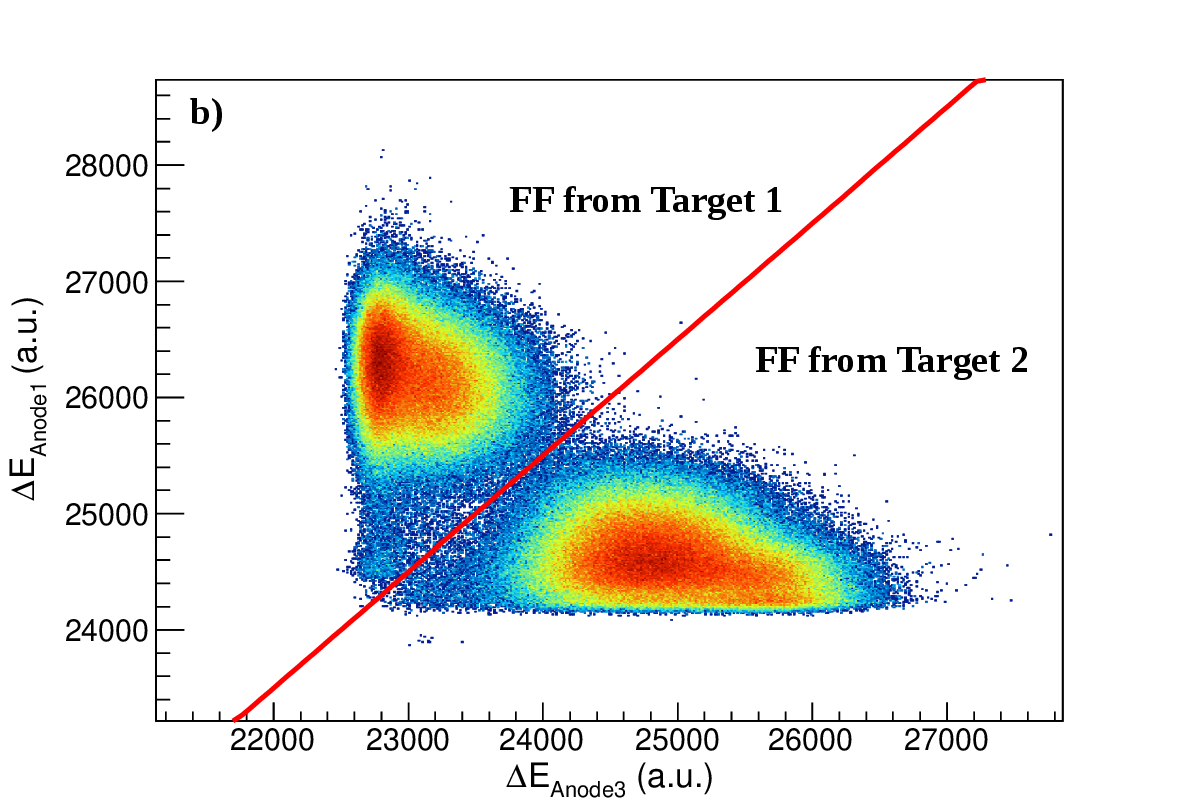}
\includegraphics[width=0.49\textwidth]{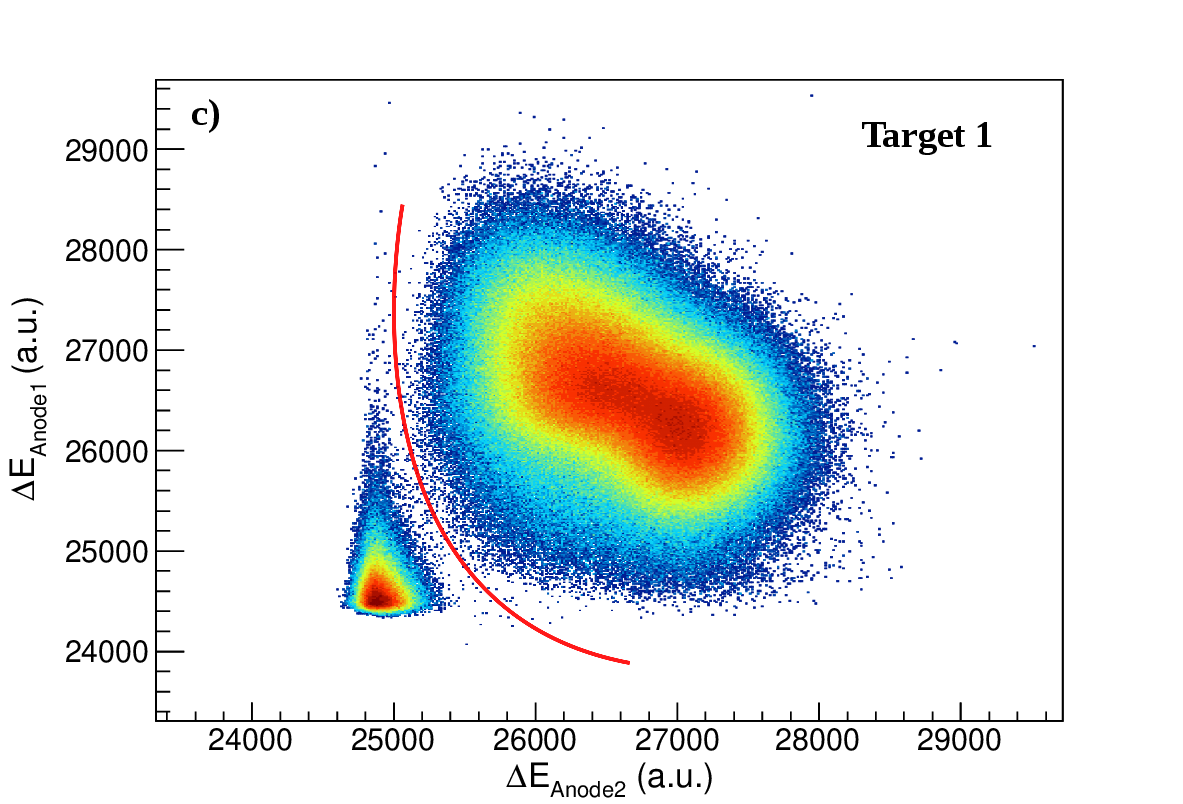}
\includegraphics[width=0.49\textwidth]{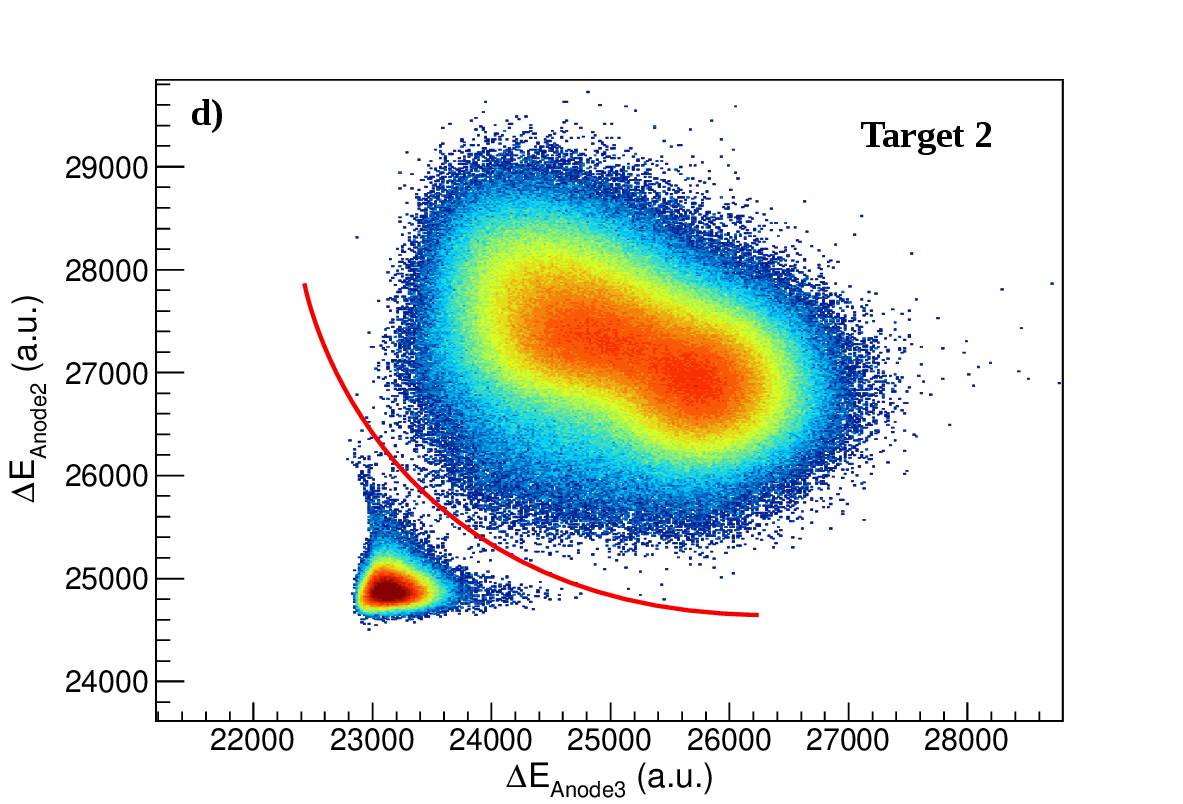}
\caption{Selection of fission fragments from anode signals. a) Energy loss correlation detected in the anodes of two PS-PPACs surrounding the target 1. Fission fragments (FF) are well separated from alpha particles. Fission fragments from target 2, crossing PS-PPAC 2 and arriving to PS-PPAC 1 are also detected. b) Energy loss correlation of fission fragments between PS-PPAC 1 and PS-PPAC 3 in order to separate fission fragments coming from the target 1 and from the target 2. c) and d) Final identification of fission fragments from corresponding targets. The red lines represents the discrimination between alpha particles and fission fragments applied in the analysis.}
\label{fig_TarSelection}
\end{figure*}

Three detectors and the two targets are placed in a dedicated reaction chamber, as shown in Fig.~\ref{fig_chamber}, equipped with two 125 $\mu$m-thick Kapton windows of 153 mm diameter on the front and back sides of the chamber along the beam axis to reduce the neutron beam interation. The entire chamber is filled with 7 mbar of iC$_4$H$_{10}$ gas, with a continuous 8 sccm flow guided through the gaps between the anodes and the cathodes. 

The anodes are polarized at +540 V while the cathodes are grounded. A home-made current-sensitive preamplifier filters the HV and provides a fast output signal with 6 ns rise time from the anode. This signal is split in two: the first provides a measurement of the energy loss of the fragment by integrating the signal over 100 ns, while the second is sent to a constant fraction discriminator (CFD) to generate a logic gate. The logic OR of the three anode gates defines the Start for the neutron time-of-flight in a reversed configuration, where the Stop is provided by a signal synchronized with the radiofrequency of the LINAC cavities and delayed to arrive after the Start. Both, Start and Stop signals are sent to a Time to Amplitude Converter (TAC). 

Each cathode plate provides two signals, one from each edge of the delay line. They are amplified through a charge-sensitive preamplifier and sent to a CFD. The logic gate of each CFD is sent to a TAC. The right/top signal provides the Start, and the left/bottom signal, previously delayed by 500 ns, provides the Stop. 

Home-made NUMEXO2 digital electronics~\cite{numexo} read out both anode and TAC outputs. Each NUMEXO2 channel operates in trigger-less mode with an associated 100MHz timestamp, which is used to correlate the events during offline analysis~\cite{merge}.   

\section{Data Analysis}
\label{sec_analysis}

\subsection{Selection of Fission Fragments}
\label{sec_selection}

Fission reactions from the $^{238}$U targets are tagged by the simultaneous detection of two fission fragments in the PS-PPACs surrounding each target. Those fission events must be separated from various sources of contamination in order to obtain a counting rate that is representative of the neutron flux.

The energy loss of the fragments is used to distinguish fission fragments from alpha particles, which are also detected by the PS-PPACs but exhibit lower energy loss. Additionally, when the emission angle of the fragments is close to 0 deg, one single light fragment may traverse two detectors producing coincident signals.

The correlation between the energy loss in both PS-PPACs surrounding a given target allows for the identification of different detected ions. Figure~\ref{fig_TarSelection}a) shows the energy loss measured in the anode of PS-PPAC 2 versus that in the anode of PS-PPAC 1. Alpha particles are clearly separated from fission fragments. Fission fragments coming from the neighboring target are identified by a similar energy loss in PS-PPAC 2 but lower energy in PS-PPAC 1, compared to the fragments of interest. 

The assignment of the fission fragments source is performed using the correlation between the energy loss in PS-PPAC 3 and PS-PPAC 1, as shown in Fig.~\ref{fig_TarSelection}b). The red line indicates the separation limit defined in the analysis for fragments originating from targets 1 and 2. Fragments from the neighboring target are thus excluded from the analysis to avoid double counting. 

Figures~\ref{fig_TarSelection}c) and~\ref{fig_TarSelection}d) show the final identification of fission fragments from target 1 and 2, respectively. The red curve indicates the cut applied in the analysis to separate alpha particles from fission fragments. 

\subsection{Calculation of the Neutron Energy}
\label{sec_neutron_energy}

The neutron energy ($E_n$) is determined by the time-of-flight (ToF) between the $^9$Be-converter target and the PS-PPAC that provides the faster signal on an event-by-event basis. 
The neutron ToF ($ToF_n$) is corrected by subtracting the time-of-flight of the fragments from the $^{238}$U target to the corresponding PS-PPAC. This correction assumes an average light-fragment velocity of 1.4 cm/ns ---extracted from Ref.~\cite{vives}--- and a travel distance from 1.7 cm to 3.4 cm, depending on the fragment emission angle, as presented in Sec.~\ref{sec_geometrical_efficiency}. A global time offset is included to account for the various time delays introduced by the electronic system. The neutron energy is calculated as:

\begin{equation}
E_n = m_nc^2(\gamma-1); \\
\gamma = \frac{1}{\sqrt{1-(\frac{D}{c(ToF_n)})^2}}.
\end{equation}
where $m_n$ is the mass of the neutron, $c$ is the speed of light, and $D$ is the distance between the converter target and the $^{238}$U target. 

\begin{figure}[t!]
\includegraphics[width=0.49\textwidth]{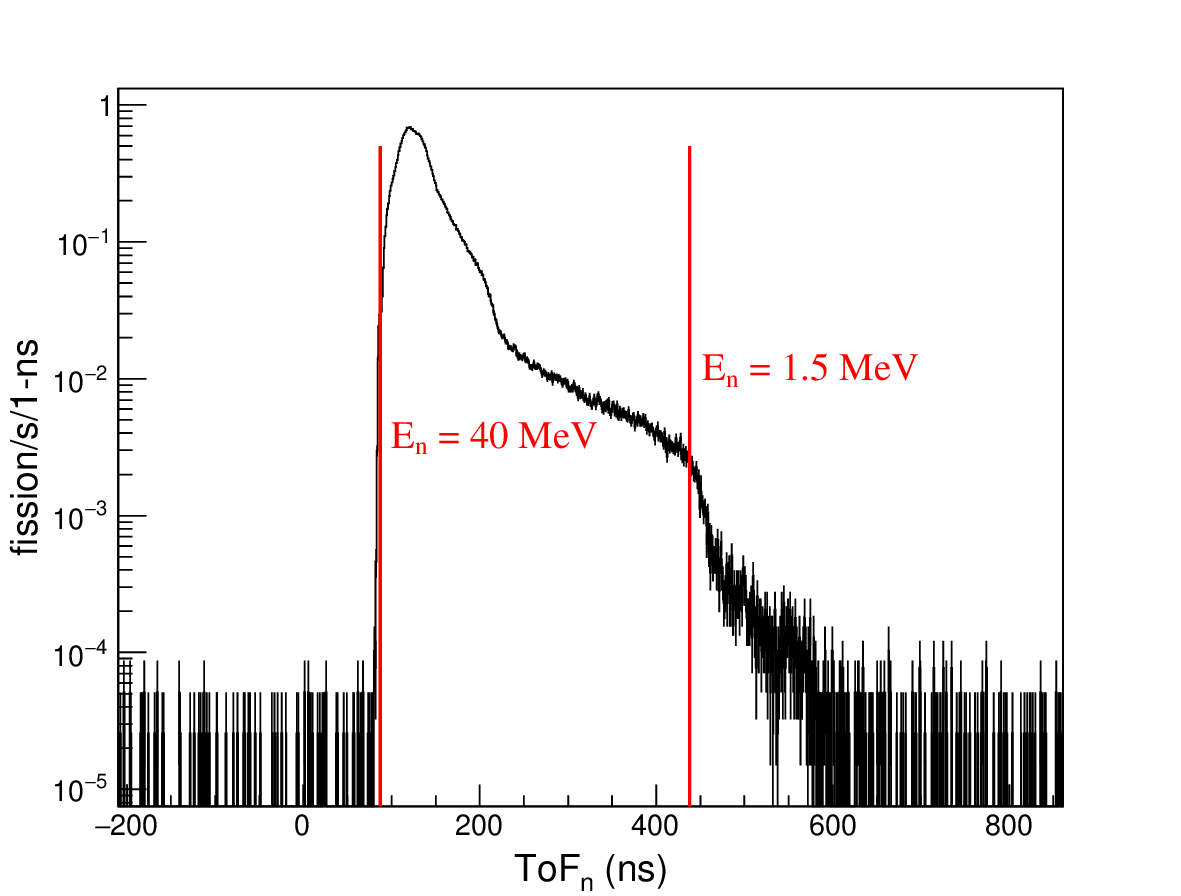}
\caption{Rate of detected fission fragments as a function of the neutron Time of Flight. Vertical lines indicates the $ToF_n$ corresponding to neutron energies of 1.5~MeV and 40~MeV.}
\label{fig_nTOF}
\end{figure}

Figure~\ref{fig_nTOF} shows the distribution of fission fragments detected in the setup as a function of $ToF_n$. The $ToF_n$ values corresponding to neutron energies of 1.5 MeV and 40 MeV are indicated by vertical lines. A low, constant background as a function of $ToF_n$ is observed in the spectrum for $ToF_n$ below 80 ns and above 600 ns. This background is understood to originate from fission reactions induced by thermalized neutrons interacting with the small amount of $^{235}$U present in the target. It represents a fraction of $(1.40\pm0.16)\times10^{-4}$ of the total number of detected fission reactions. Given its small contribution, this background is negligible compared to other sources of uncertainty, as will be discussed later, and it is therefore not considered in the neutron flux calculation. 

\begin{figure}[t!]
\includegraphics[width=0.49\textwidth]{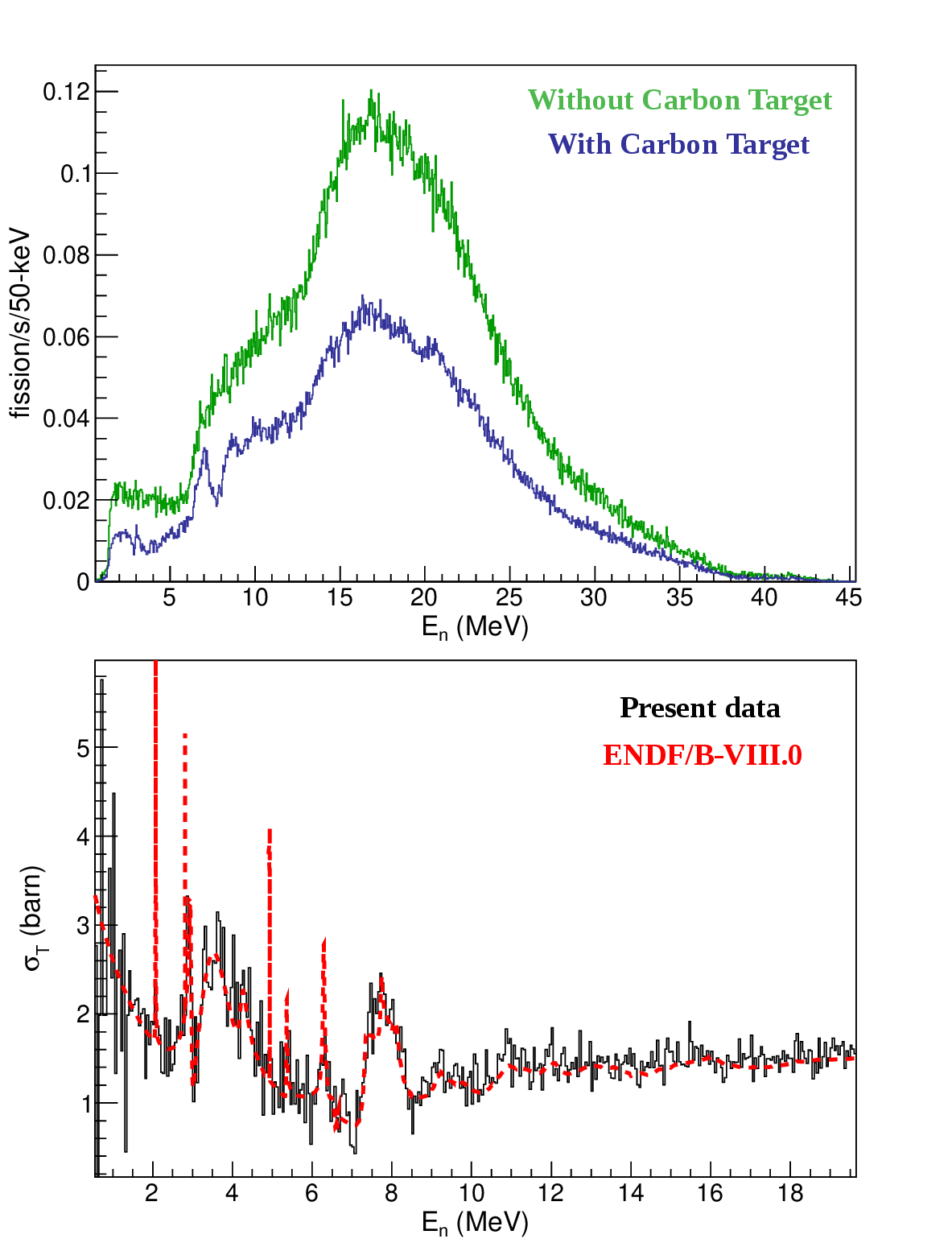}
\caption{Neutron energy calibration from Time-of-Flight. (Top) $^{238}$U(n,f) fission fragments production as a function of the neutron energy from a neutron beam impinging directly into the U target (green) and from an attenuated neutron beam passing through 4-cm Carbon target before reaching the U target (blue). (Bottom) Total absorption cross section of $^{12}$C(n,tot) as a function of the neutron energy from present data (black) compared with reference data from ENDF/B-VIII.0 data base (red).}
\label{fig_transmission}
\end{figure}

No $\gamma$-flash is present at NFS in sufficient intensity to detect fission fragments from photofission~\cite{NFS3}; therefore, this timing reference is not available. Instead, the global time offset is determined using the well-know $^{12}$C(n,tot) resonances between 2 MeV and 10 MeV in neutron energy as an absolute reference. The $^{238}$U(n,f) fission-fragment production is measured with the neutron beam directly impinging on the uranium target, and with the beam passing through a 4-cm-thick carbon target placed in front of the uranium target. Fig.~\ref{fig_transmission} (top) shows both measurements, normalized to the same integrated deuteron-beam intensity. The total absorption cross section is determined from the ratio of these two measurements, applying the transmission technique as discussed in~\cite{transmission}. Figure~\ref{fig_transmission} (bottom) presents the total absorption cross section of $^{12}C(n,tot)$ where the present data (in black) is compared with reference data from ENDF/B-VIII.0~\cite{ENDF} (in red) as a function of the neutron energy.

\subsection{Geometrical Detection Efficiency}
\label{sec_geometrical_efficiency}

The intrinsic efficiency of the PS-PPACs for detecting fission fragments is assumed to be 1. However, when the emission angle of the fission fragments reaches a critical value, the increased effective thickness of the target, backing, and Mylar foils may cause the fission fragments to be stopped either inside the target, the backing, or the detector itself ---before reaching the last gas gap required to produce a signal in the last cathode--- thus reducing the overall detection efficiency of the setup.
 
\begin{figure}[t!]
\includegraphics[width=0.49\textwidth]{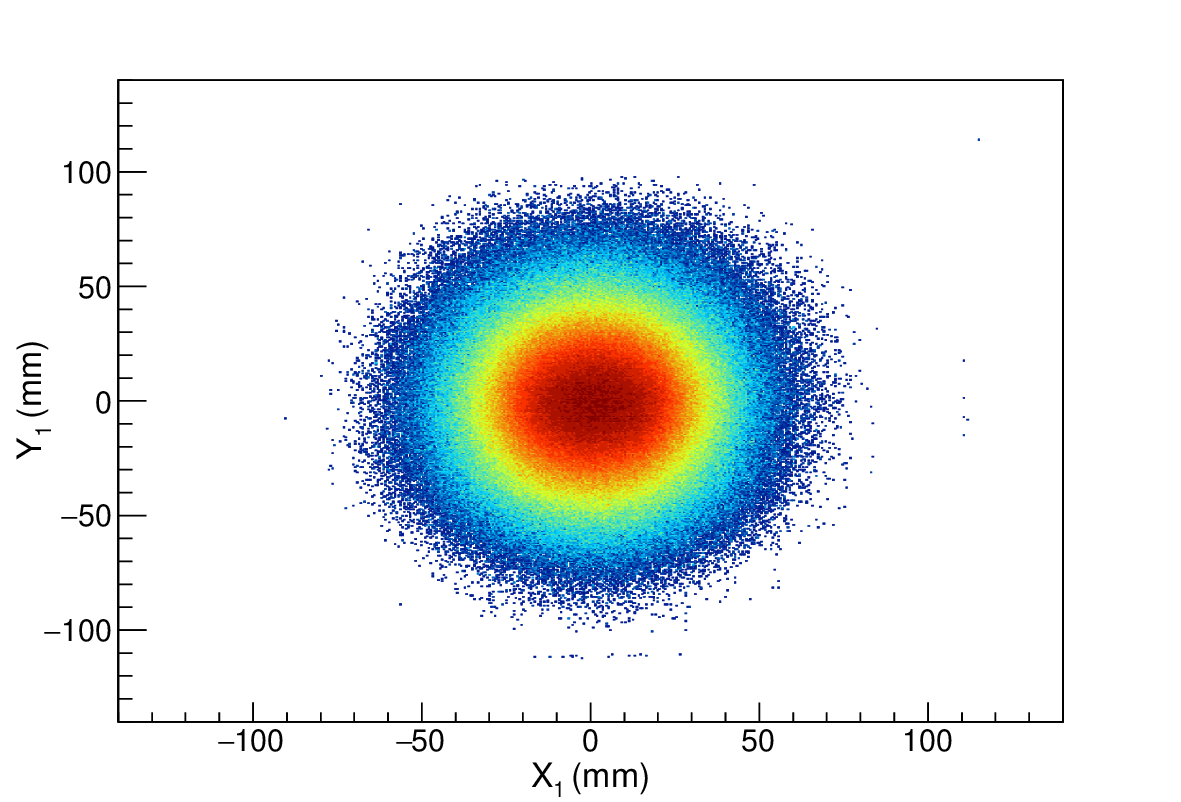}
\caption{Position of the fission fragments from $^{238}$U(n,f) coming from target 1, detected in PS-PPAC 1.}
\label{fig_PPAC_XY}
\end{figure}

Figure~\ref{fig_PPAC_XY} shows the hit position ($X,Y$) in the PS-PPAC 1 of fission fragments coming from target 1, selected using the energy-loss correlation described in Sec.~\ref{sec_selection}. The effect of geometrical efficiency is observed at radial positions beyond 50 mm from the center, where the number of detected fragments decreases rapidly. 

The emission angle of fission fragments is used to experimentally determine the geometrical efficiency of the setup. The polar angle ($\theta_i$) ---i.e., the angle between the fission fragments and the neutron beam--- is reconstructed event by event from the position of both coincident fission fragments ($X_{i},Y_{i}$) and ($X_{i+1},Y_{i+1}$), detected in the two PS-PPACs surrounding the target $i$, under the assumption that both fragments are emitted back-to-back\footnote{This assumption is valid up to 40 MeV within an uncertainty of $\pm 0.007$ in $\cos(\theta_i)$, as determined following the method presented in Ref.~\cite{Tarrio1}.}:
\begin {equation}
\tan{\theta_{i}} = \frac{\sqrt{(X_{i+1}-X_{i})^2+(Y_{i+1}-Y_{i})^2}}{D_i^{i+1}},
\end {equation}
where $D_i^{i+1}$ is the distance between the detectors $i$ and $i+1$ along the axis perpendicular to the detector planes.

\begin{figure}[h!]
\includegraphics[width=0.49\textwidth]{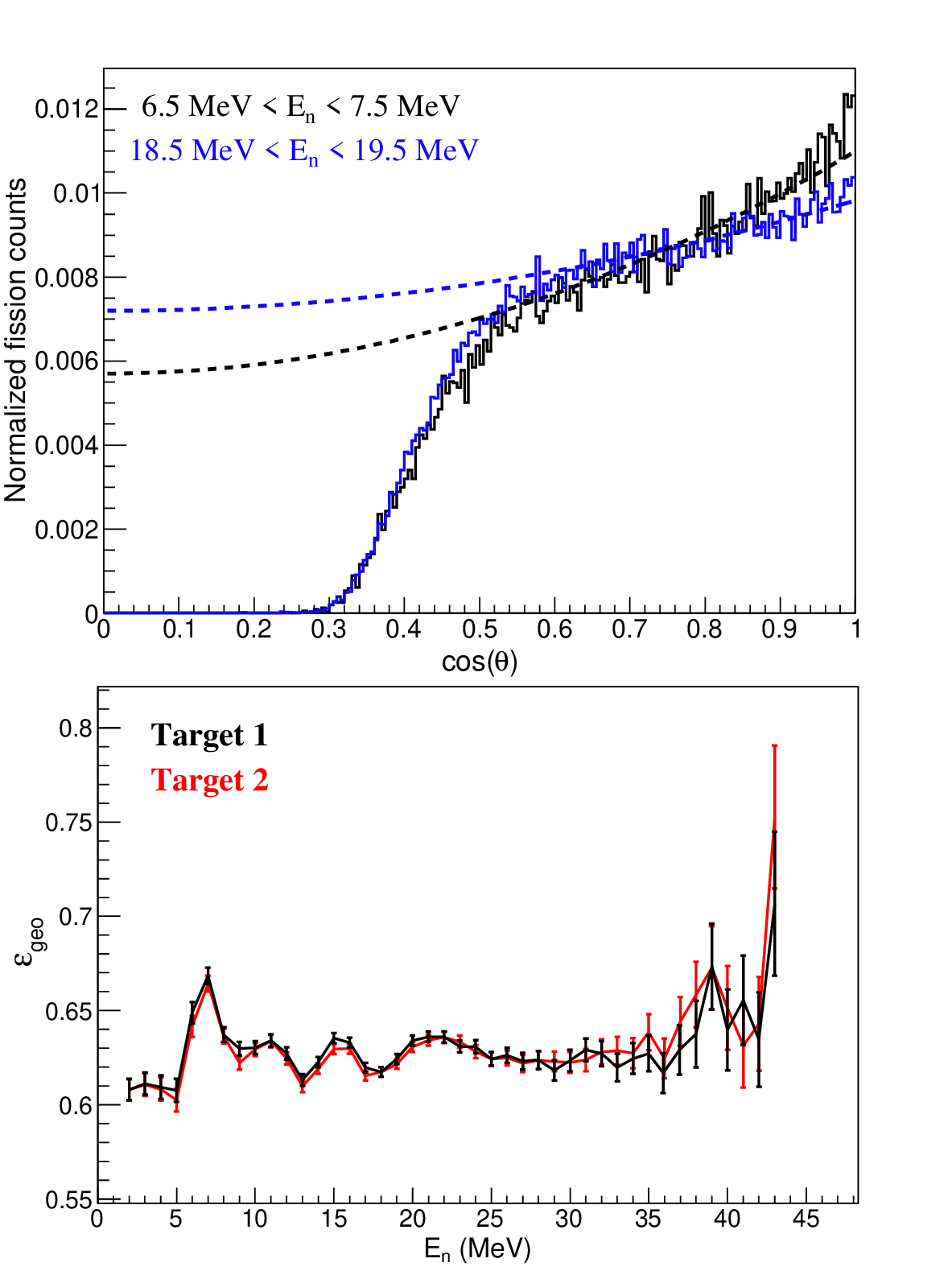}
\caption{(Top) Angular distribution of fission fragments detected in the current setup (continuous line) compared to the expected distribution considering angular anisotropy (dashed lines). Two distributions from different neutron energy ranges are presented, $E_n = 7\pm0.5$~MeV in black and $E_n = 19\pm0.5$~MeV in blue. (Bottom) Detection geometrical efficiency as a function of the neutron energy. Coincident fission fragments from both targets are presented.}
\label{fig_Anisotropy}
\end{figure}

Figure~\ref{fig_Anisotropy} (top) shows the polar angular distribution of fission fragments from target 1 as a function of $\cos(\theta)$ for two neutron energy ranges: $E_n = 7\pm0.5$~MeV (black) and $E_n = 19\pm0.5$~MeV (blue). Both distributions are normalized to the same integral. A drop in detection efficiency is evidenced at $\cos(\theta) < 0.6$, where the number of detected fragments decreases to zero. Additionally, it is well know that $^{238}$U(n,f) exhibits positive angular anisotropy, favoring lower emission angles with respect to the beam axis\footnote{In this setup, the beam axis defines the reference axis of the polar angle, perpendicular to the detector planes.}~\cite{Vorobyev1,Vorobyev2}. This anisotropy, observed at $\cos(\theta)> 0.6$, varies with the neutron energy and modifies the detection efficiency.

The geometrical efficiency is determined, as a function of the neutron energy, by taking the ratio between the number of detected fission events and the number expected from anisotropic angular distribution described by a second-order Legendre polynomial:
\begin{equation}
W(\cos(\theta)) = A\left[1+\left(\frac{W(0^\circ)}{W(90^\circ)}-1\right)\cos^2(\theta)\right]
\end{equation}
where $\frac{W(0^\circ)}{W(90^\circ)}$ is the anisotropy factor, taken from Ref.~\cite{Vorobyev1,Vorobyev2} experimentally measured, and $A$ is a normalization factor fitted to the current data in the region $\cos(\theta)$ between 0.6 and 1. The resulting angular distributions from the Legendre polynomials are represented in Fig.~\ref{fig_Anisotropy} (top) as dashed lines for $E_n = 7\pm0.5$~MeV (black) and $E_n = 19\pm0.5$~MeV (blue).

Figure~\ref{fig_Anisotropy} (bottom) presents the geometrical efficiency as a function of neutron energy. Data from both targets are included  and they show similar values. A maximum of geometrical efficiency appears around $E_n = 7$ MeV, corresponding to a maximum in angular anisotropy. The uncertainties shown are obtained as the quadratic sum of statistical uncertainties of present data, the uncertainties in the reference anisotropy measurement, and the uncertainty of the fit. In the region above 35 MeV, low statistics prevent an accurate determination of the detection efficiency. In this region, an additional systematic uncertainty of 10\% is added to the final fission rate.    

\section{Results}
\label{sec_results}
\subsection{Neutron Beam Profile}
\label{sec_profile}

\begin{figure}[!]
\begin{center}
\includegraphics[width=0.4\textwidth]{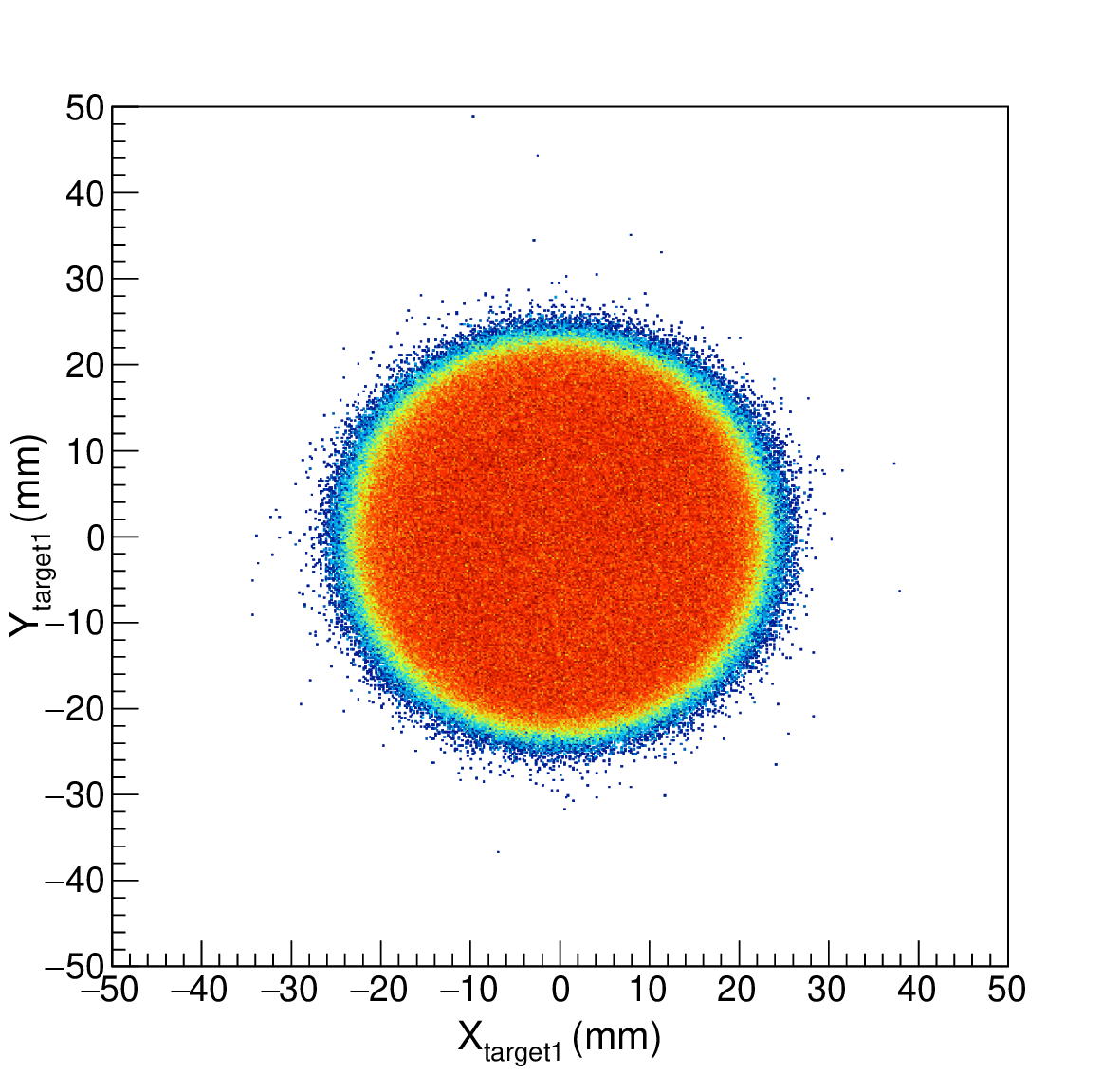}
\end{center}
\caption{Neutron beam profile at the first $^{238}$U target determined by the fission reaction point reconstruction.}
\label{fig_BeamSpot}
\end{figure}

The fission reaction point at the uranium target ($X_{target_i},Y_{target_i}$) is reconstructed on a event-by-event basis by tracking the position of the two coincident fission fragments in the PS-PPACs surrounding the target: 
\begin{equation}
\begin{split}
X_{target_i}= X_i+\frac{D_{i}^{target_i}}{D_i^{i+1}}(X_{i+1}-X_i), \\
Y_{target_i}= Y_i+\frac{D_{i}^{target_i}}{D_i^{i+1}}(Y_{i+1}-Y_i),
\end{split}
\end{equation}
where $D_{i}^{target_i}$ is the distance between the PS-PPAC $i$ and the target $i$, and $D_i^{i+1}$ is the distance between both PS-PPACs $i$ and $i+1$, both along the axis perpendicular to the detector planes. The spatial segmentation of the detectors provides a resolution in the reconstruction of the reaction point of $\sigma=0.43$ mm. 

The neutron beam profile is smaller than the diameter of the uranium target; thus, the spatial distribution of reconstructed fission reactions at the target directly reflects the neutron beam profile. Figure~\ref{fig_BeamSpot} shows the measured beam profile at the first uranium target. The beam spot has a diameter of $D_1 = 44.4\pm0.4$ mm, with a fall-off region of $4.9\pm0.4$ mm, measured between 10\% and 90\% of the maximum number of detected events per surface unit.

\begin{figure}[!]
\includegraphics[width=0.49\textwidth]{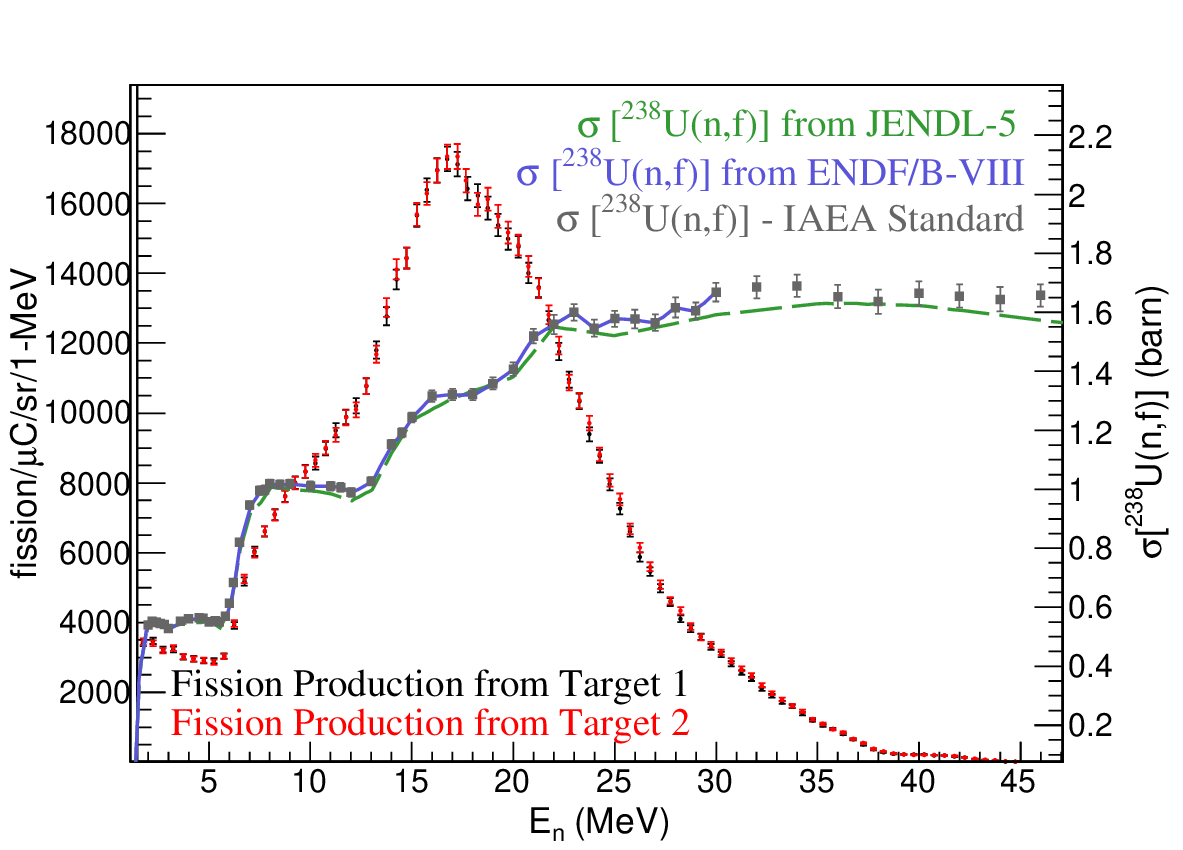}
\caption{$^{238}$U(n,f) production in target 1 (black) and target 2 (red) as a function of the neutron energy, with the scale in the left vertical axis; and the $^{238}$U(n,f) cross section from nuclear data library JENDL-5 (green dashed line) and ENDF/B-VIII (solid blue line), and the IAEA standard (gray squares), with the scale in the right vertical axis.}
\label{fig_FissionProd}
\end{figure}

\begin{figure}[!]
\includegraphics[width=0.49\textwidth]{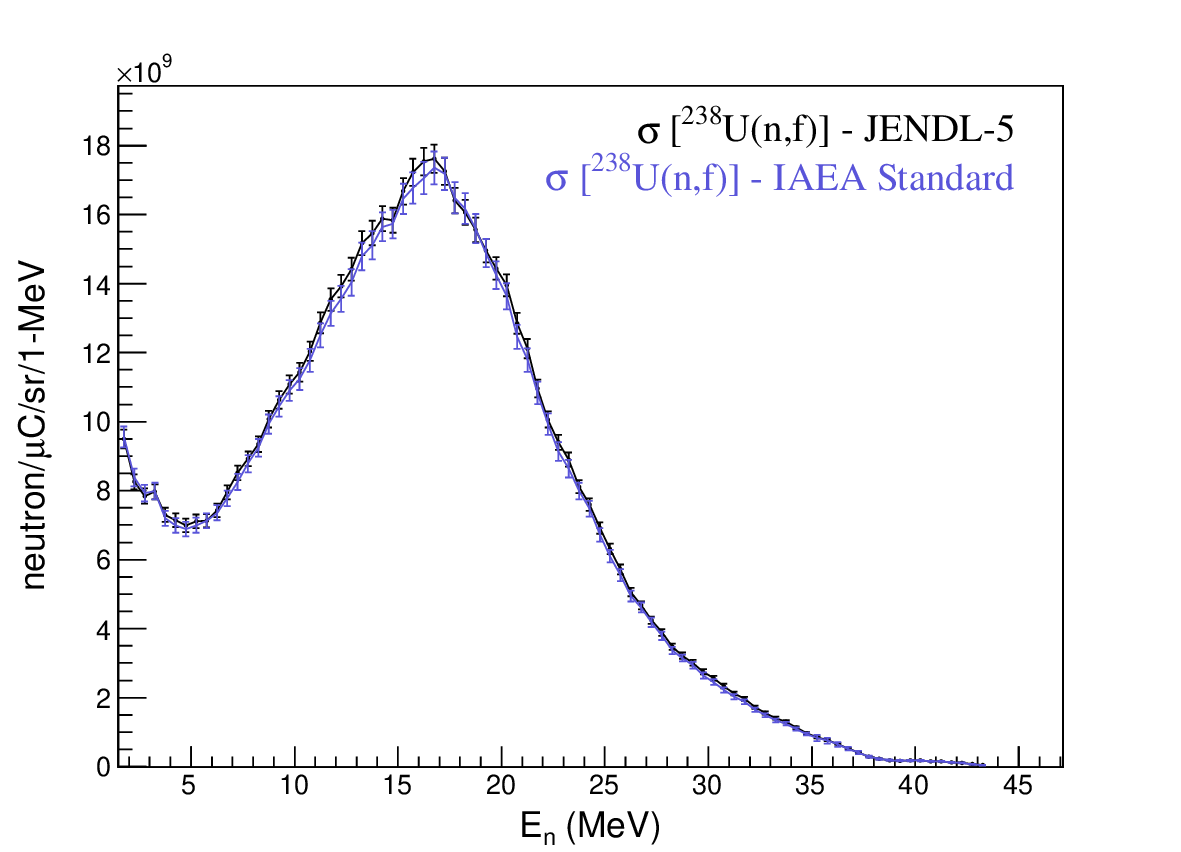}
\caption{Neutron rate production from $^{238}$U(n,f) counting rate, corrected by the $^{238}$U(n,f) cross section obtained from JENDL-5 (black) and from IAEA (blue).}
\label{fig_NeutronProd2}
\end{figure}

The beam profile is shaped by the opening of the neutron-beam collimator, with a diameter of $D_c = 25.55$~mm, placed $\Delta L=3052.9$~mm upstream of the uranium target. From both diameters, the beam divergence can be calculated as: 
\begin{equation}
\theta = \arctan\frac{D_1-D_c}{2\Delta L},
\end{equation}
which yields a beam divergence of $\theta = 3.08\pm 0.07$ mrad.

\subsection{Neutron Beam Flux}
\label{sec_flux}

\begin{figure}[!]
\includegraphics[width=0.49\textwidth]{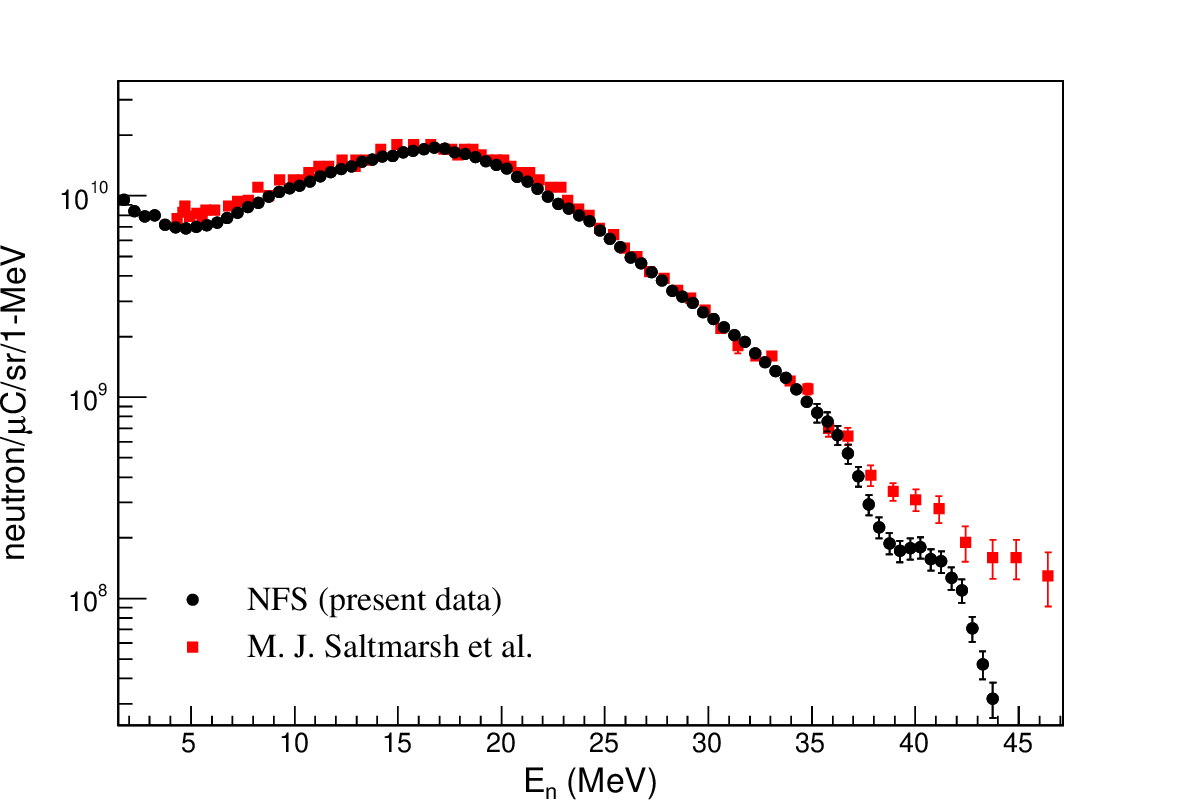}
\caption{Neutron beam flux at NFS from a primary deuteron beam at 40 MeV impinging into a 8 mm-thickness Be converter target (black dots) compared with previous measurement of the neutron production from 40-MeV deuteron beam impigning into a 6.3 mm-thickness Be target (red squares), from Ref.~\cite{Saltmarsh}.}
\label{fig_NeutronProd}
\end{figure}

The neutron-beam flux is determined from the rate of fission reactions induced in the $^{238}$U targets. The number of fission fragments detected in coincidence in two PS-PPACs is corrected for the detection efficiency, presented in Fig.~\ref{fig_Anisotropy} (bottom), as a function of the neutron energy. This count is then normalized by the solid angle of the neutron beam at the uranium-target position, $\Omega = (28.2\pm0.5)\cdot 10^{-6}$ sr, which is obtained from the measured beam profile shown in Fig.~\ref{fig_BeamSpot}, and by the intensity of the primary deuteron beam, $I_{deuteron}=7.60\pm 0.05$ $\mu$A, measured regularly using a Faraday cup placed 25.3 cm upstream of the Be converter target. 

The resulting rate of fission reactions detected in both $^{238}$U targets, expressed per second, per $\mu$A of deuteron-beam current, per steradian, and per 1-MeV neutron-energy steps, is presented in Fig.~\ref{fig_FissionProd} as a function of the neutron energy. Both $^{238}$U targets exhibit similar counting rates, as they have comparable thickness and no significant beam attenuation is expected between them.
 
Figure~\ref{fig_FissionProd} also displays the $^{238}$U(n,f) cross section from two evaluated nuclear data libraries: JENDL-5~\cite{JENDL} (dashed green line) and ENDF/B-VIII~\cite{ENDF} (solid blue line), as well as the standard adopted by the IAEA~\cite{IAEA} (gray squares). While ENDF/B-VIII data extend up to 30 MeV, JENDL-5 and IAEA data cover the full neutron energy range explored in this work. Significant differences are observed between the JENDL-5 and IAEA datasets.

Fig.~\ref{fig_NeutronProd2} shows the deduced neutron flux as a function of the neutron energy, calculated from the fission rate in the $^{238}$U target 1. The data are corrected for the target thickness ($298\pm5 {\mu}g/cm^2$) and the $^{238}$U(n,f) cross section from JENDL-5 (black) and from IAEA (blue). Both datasets provide consistent neutron flux values within their respective uncertainties. Therefore, the IAEA standard is adopted for the final analysis.  

The final neutron flux, produced by reactions between a 40-MeV primary deuteron beam and an 8-mm-thick beryllium converter target, is shown in Fig.~\ref{fig_NeutronProd} (logarithmic scale), expresed in units of neutrons per second, per $\mu$A of the primary beam current, per steradian, and per 1-MeV-neutron energy interval, as a function of neutron energy from 1.5 MeV to 44 MeV. A flux above $10^8$ neutrons/$\mu$C/sr/1-MeV is observed from $E_n=1.5$ MeV to $E_n=42$ MeV, and above $10^9$ neutrons/$\mu$C/sr/1-MeV between $E_n=1.5$ MeV and $E_n=35$ MeV. The maximum neutron production occurs at $E_n = 17$ MeV with a yield of $(17.3\pm0.5)\cdot10^9$ neutrons/$\mu$C/sr/1-MeV. An interesting feature is observed at high energies: neutrons with energies exceeding the 40-MeV deuteron beam energy are detected. This indicates the presence of neutron-emitting reactions with a positive Q-value, as discussed in Ref.~\cite{Brede,Iwamoto}.

Present data from NFS (black dots) are compared with a previous measurement reported in Ref.~\cite{Saltmarsh} (red squares), in which neutrons were produced from rections between a 40-MeV deuteron beam and 6.3 mm-thick beryllium target. The overall good agreement between both datasets demonstrates the high quality of the current results and validates this experimental technique employed. The differences observed at higher neutron energies are attributed to the limited energy resolution in the previous measurement constrained by a short flight path of only 3 meters.
 
\subsection{Estimation of Thermal Neutrons from ToF-uncorrelated Fission Background}
\label{sec_Thermal}

As discussed in Sec.~\ref{sec_neutron_energy}, the amount of fission reactions not correlated in time with the neutron beam introduces a constant background in the $ToF_{n}$ distribution. This background is mainly attributed to fission reactions induced by thermal neutrons on the 0.72\% fraction of $^{235}$U present in the target. Consequently, the amount of thermal neutrons in the experimental area can be upper-bounded by comparing the rate of these $ToF_{n}$-uncorrelated fission reactions to that of the $ToF_{n}$-correlated ones, resulting in a ratio of $\frac{N_{fission}^{background}}{N_{fission}^{corr}}= (1.40\pm0.16)\times10^{-4}$. Using the thermal-neutron-induced fission cross section of $^{235}$U, $\sigma_{th}^{235} = 586.1$ barns, and the average neutron-induced fission cross section of $^{238}$U over the studied energy range from 1.5 to 44~MeV, $\langle\sigma^{238}\rangle= 1.17$ barns ---both values are from the IAEA standard--- the upper limit on the relative amount of thermal neutrons with respect to fast neutrons ($\frac{n_{th}}{n_{fast}}$) is calculated as: 
\begin{equation}
\frac{n_{th}}{n_{fast}} \leq \frac{N_{fission}^{background}}{N_{fission}^{corr}}\times\frac{\langle\sigma^{238}\rangle}{\sigma_{th}^{235}}\times\frac{1}{R_{target}^{235/238}},
\end{equation}
where $R_{target}^{235/238}=0.0072$ is the atomic ratio of $^{235}$U over $^{238}$U in the target. 

This calculation yields an upper limit on thermal neutron production of $\frac{n_{th}}{n_{fast}}\leq(3.85\pm 0.43)\times10^{-5}$.

\section{Conclusions}
\label{sec_conclusions}

Fission of $^{238}$U, coupled with a high-selectivity instrument such as PS-PPACs, provides a powerful tool for measuring neutron flux in fast-neutron energy region. This technique was applied to the NFS neutron beam, produced via reactions between a 40-MeV deuteron beam and a rotating $^{9}$Be converter. The maximum neutron flux of $(17.3\pm0.5)\cdot10^9$ neutrons/$\mu$C/sr/1-MeV was measured at $E_n=17$~MeV. The neutron flux remains above $10^9$ neutrons/$\mu$C/sr/1-MeV throughout the energy range from $E_n=1.5$ MeV to $E_n=35$ MeV. These results are in good agreement with previous simulations of the facility~\cite{NFS2}. Additionally, the position sensitivity of the detectors enabled reconstruction of the beam profile, revealing a fall-off region of $4.9\pm0.4$ mm. The low background level indicates a suppressed production of thermal neutron, with an upper limit of $(3.85\pm 0.43)\times10^{-5}$ relative to the total fast neutrons population.

\end{document}